\documentclass[12pt]{article}
\usepackage[utf8]{inputenc}
\usepackage[margin=0.75in]{geometry}
\usepackage{graphicx}
\usepackage{tabularx}
\usepackage{dcolumn}% Align table columns on decimal point
\newcolumntype{C}{>{\centering\arraybackslash}X}
\newcolumntype{d}[1]{D{.}{.}{#1}}

\usepackage[caption=false]{subfig}
\usepackage[square,numbers]{natbib}
\setcitestyle{number}

\title{
Technique for rapid mass determination of airborne micro-particles based on release and recapture from an optical dipole force trap}
\begin{document}

%\maketitle
{\LARGE Technique for rapid mass determination of airborne micro-particles based on release and recapture from an optical dipole force trap}

\vspace{5mm}

%\author{Gehrig Carlse, Kevin B. Borsos, Hermina C. Beica, Thomas Vacheresse, Alex Pouliot, Jorge Perez-Garcia, Andrejs Vorozcovs, Boris Barron, Shira Jackson, Louis Marmet, A. Kumarakrishnan}%
% \email{akumar@yorku.ca}
Gehrig Carlse, Kevin B. Borsos, Hermina C. Beica, Thomas Vacheresse, Alex Pouliot, Jorge Perez-Garcia, Andrejs Vorozcovs, Boris Barron, Shira Jackson, Louis Marmet, A. Kumarakrishnan

\vspace{2mm}

%\affiliation{%
%Department of Physics and Astronomy, York University, Toronto ON, Canada M3J 1P3\\

 
%\begin{description}
%\end{description}

%\maketitle

%\tableofcontents
\textit{Department of Physics and Astronomy, York University, Toronto ON, Canada M3J 1P3}
\section*{Abstract}
We describe a new method for the rapid determination of the mass of particles confined in a free-space optical dipole-force trap.  The technique relies on direct imaging of drop-and-restore experiments without the need for a vacuum environment.
 In these experiments, the trapping light is rapidly shuttered with an acousto-optic modulator causing the particle to be released from and subsequently recaptured by the trapping force.  The trajectories of both the falls and restorations, imaged using a high-speed CMOS sensor, are combined to determine the particle mass.  
 We corroborate these measurements using an analysis of position autocorrelation functions of the trapped particles.  We report a statistical uncertainty of less than 2\% for masses on the order of $5\times10^{-14}$~kg using a data acquisition time of approximately 90 seconds.

\section{Introduction}

The development of optical dipole-force (ODF) laser traps to confine dielectric particles \cite{ashkin1974stability,ashkin1975optical,ashkin1976optical,ashkin1970acceleration,ashkin1986observation} has led to wide-ranging applications across many fields of research.  Notable examples include the development of far-off resonance traps (FORTs) for confining atoms \cite{wieman1999,chu1999ODF}, the use of optical tweezers to generate three-dimensional optical crystals \cite{raithel1,raithel2}, and the application of ODF traps for the manipulation of biological molecules \cite{ashkin1990force,ashkin1987bacteria}, measurements of bond strengths \cite{chu1995stretching}, and protein synthesis \cite{fazal2015direct,cecconi2008protein}.

The progression of this field has led to powerful experiments investigating the diffusive kinematics of single particles trapped in liquids or free space.  The pioneering experiments in references \cite{huang2011direct,li2010measurement,li2013brownian,lukic2007motion,lukic2005direct} have enabled investigation of the timescales on which diffusive Brownian motion transitions to ballistic motion.  
Other lines of inquiry have focused on particle kinematics to study the properties of the trap itself \cite{burnham2010parameter,leonardo2007parametric,viana2007towards}, the color of the stochastic force associated with Brownian motion \cite{franosch2011resonances}, the development of precise force sensors \cite{hebestreit2018sensing}, determinations of fluid viscocity \cite{guzman2008situ,grimm2012high}, and measurements of the polarizability of trapped particles \cite{dielectric}.
Progress in these areas has focused on improving detection bandwidth \cite{chavez2008development} and spatial resolution \cite{staforelli2010superresolution} to investigate smaller time and length scales.
These experiments have employed complementary techniques, such as analyses of power spectral densities \cite{berg2004power,berg2006power}, and position and velocity autocorrelation functions \cite{li2013brownian,huang2011direct}.

Recently, there has also been widespread interest in employing optical tweezers to perform precise mass measurements of trapped particles \cite{nanoricci2019accurate,blakemore2019precision,western}.  Table \ref{tab:masssummary} shows a representative compilation of such tweezers-related measurements. The most sensitive and accurate mass measurement involving optical tweezers has been obtained using underdamped ODF traps operated in a vacuum environment \cite{nanoricci2019accurate}.  In reference \cite{nanoricci2019accurate}, the trapped particle was driven using an alternating electric field and the mass was determined by fitting to the power spectral density of the motion.  This technique has been successful in characterizing masses 
%of $\sim 4\times 10^{-18}$~kg, 
on the femtogram scale with a precision of 0.25\%.  Other examples of mass determinations in the range of $10^{-10} - 10^{-15}$~kg involve photophoretic traps \cite{chen2018drop} and have achieved precision at the level of a few percent \cite{lin2017measurement,bera2016simultaneous}.  

In this paper, we show that the use of video microscopy to track the release and recapture of particles held in a free-space
%single-beam gradient or single beam-gradient
single-beam gradient trap results in a simple technique for the rapid and precise determination of the particle’s mass.  
This technique does not require a vacuum environment or electro-mechanical feedback systems.
Additionally, it is demonstrated here using modest laser powers and a small field of view.
The precise timing required for the release and recapture of trapped particles is enabled by amplitude modulation using an acousto-optic modulator (AOM).  As a result, we combine the advantages of tight confinement in an ODF trap and the capacity to observe unconstrained particle kinematics sensitively as in the drop-tower studies of \cite{blum2006measurement}.  Building on techniques to study the ballistic expansion of ultracold atomic samples \cite{carlse2020technique}, we track the centroid of particles dropped in free space to infer the damping rate and analyze the trajectory of the recaptured particle to determine the particle mass.  These measurements are corroborated by separate studies of the position autocorrelation function (PACF).  We show that masses on the order of $10^{-14}$~kg associated with resinous particles with diameters of a few micrometers can be determined with a statistical precision of $\sim 2\%$ in measurement times of approximately $90$~s. 

In what follows, we first describe the theoretical framework for particle kinematics and the features of the PACFs in Section 2.  Section 3 outlines the experimental set-up and Section 4 presents the main results of the paper.

\begin{table*}
\caption{\label{tab:masssummary}Summary of contemporary tweezers-based mass measurements.  
The last column indicates the statistical (Stat.) and systematic (Syst.) uncertainties.  $^{***}$Solution-based experiment. $^{**}$Photophoretic trapping experiment. $^{***}$Experiments in low-pressure vacuum environments.}
%\begin{ruledtabular}
\begin{tabular}{l c l c}
Reference & Technique & Mass (kg) & Stat. \& Syst.\\ \hline
Huang \textit{et al.} (2011) \protect{\cite{huang2011direct}}$^{*}$ & Continuous VACF analysis & $1.26\times10^{-14}$ & \textless 10\% and \textless 10\% \\ 
Bera \textit{et al.} (2016) \protect{\cite{bera2016simultaneous}}$^{**}$ & Power spectrum analysis & $9.68\times10^{-11}$ & 15\% (Stat. only) \\  
Lin \textit{et al.} (2017) \protect{\cite{lin2017measurement}}$^{**}$  & Optically-forced modulation & $9.00\times10^{-13}$ & 2\% and 6\% \\ 
Chen \textit{et al.} (2018) \protect{\cite{chen2018drop}}$^{**}$  & Dynamic power modulation & $6.3\times10^{-15}$ & Not estimated \\
Blakemore \textit{et al.} (2019) \protect{\cite{blakemore2019precision}}  & Electrostatic co-levitation  & $8.40\times10^{-15}$ & 1\% and 1.8\% \\ 
Ricci \textit{et al.} (2019) \protect{\cite{nanoricci2019accurate}}$^{***}$  & Electrostatically-driven resonance & $4.01\times10^{-18}$ & 0.25\% and 0.5\% \\ 
 \hline
 This work & Drop-and-restore & $5.58\times10^{-14}$ & 1.4\% and 13\% \\ 
\end{tabular}
%\end{ruledtabular}
\end{table*}

\section{Theory}

The stochastic motion of a particle in a fluid bath can be modelled by the Langevin equation,

\begin{equation}\label{eq:Langevin}
m\frac{\partial^2 x}{\partial t^2} + \gamma \frac{\partial x}{\partial t} = F(t)    
\end{equation}

where $m$ is the mass of the particle, $\gamma$ is the damping coefficient associated with the surrounding medium, and $F(t)$ is the stochastic force that produces Brownian motion \cite{langevin1908theorie}.

This treatment can be readily modified to include a harmonic potential due to an ODF \cite{uhlenbeck1930theory,wang1945theory}:

\begin{equation}\label{eq:Langevin2}
 \frac{\partial^2 x}{\partial t^2} + \Gamma \frac{\partial x}{\partial t} + \frac{\kappa}{m} x = A(t)   
\end{equation}
where $\Gamma = \frac{\gamma}{m}$ is the damping rate, $\kappa$ is the spring constant of the ODF trap, and the stochastic acceleration is represented as $A(t) = F(t)/m$.

Investigations into such stochastic systems have centered upon the study of the power spectral density of the motion and its Fourier transform, the PACF \cite{li2013brownian}.
The characteristic timescale on which Brownian motion transitions to ballistic motion is defined by $\tau_p = \frac{1}{\Gamma} = \frac{m}{\gamma}$, known as the momentum relaxation time.  Details of the kinematics on timescales much smaller than $\tau_p$ have been investigated by \cite{li2010measurement,lukic2007motion,lukic2005direct} in both underdamped and overdamped regimes by direct computation of correlation functions.
In addition, numerous other experiments have relied on measurements of the power spectral density to extract physical properties such as the color of the stochastic force \cite{franosch2011resonances}, the viscosity of the fluid \cite{guzman2008situ}, and the polarizability \cite{dielectric} and mass of particles \cite{nanoricci2019accurate}.

For applications based on free-space experiments, it is instructive to quantify the timescale set by $\tau_p$ to better understand the details of the particle kinematics.
For a spherical particle, Stokes' law for the damping coefficient is given by $\gamma = 6 \pi r \eta$, where $r$ is the particle radius and $\eta$ is the dynamic viscosity of the surrounding medium.
The form of Stokes' law results in a momentum relaxation time that scales as \textit{r}$^2$. 
Using the equipartition theorem and the kinetic theory of gases in which colliding particles are treated as hard spheres, it can be shown that the viscosity of a medium is described by $\eta = \frac{1}{6r_g^2}\sqrt{\frac{k_B T m_g}{\pi^3}}$, where $k_B$ is the Boltzmann constant, $T$ is the temperature, and $r_g$ and $m_g$ are the radius and mass of a gas molecule \cite{bird_stewart_lightfoot_2007}.
For nitrogen gas at $T = 300$ K, the value of $\eta$ is $\sim 17~\mu$Pa$\cdot$s, which represents a reasonable estimate for the empirical viscosity of air ($18~\mu$Pa$\cdot$s) \cite{gilchrist1913absolute}.
For a particle of radius 3 $\mu$m and mass $\sim 10^{-13}$ kg immersed in air at room temperature, this treatment
gives a momentum relaxation time of $\sim100~\mu$s.
While our experiments are designed with a temporal resolution comparable to this value of $\tau_p$, our drop-and-restore technique averages over the effects of Brownian motion by repeating the measurements on timescales much larger than $\tau_p$.  
We corroborate the resulting mass determinations using the calculation of PACFs, a complementary technique that can probe kinematics occurring on timescales of $\sim \tau_p$.

Both the PACF and the power spectral density have been calculated for Brownian motion in the overdamped, underdamped, and critically damped regimes \cite{uhlenbeck1930theory,wang1945theory,velasco1985brownian}. 
The expression for the PACF in the overdamped case, which is of interest here, is given by:

\begin{equation}\label{eq:fullPACF}
\langle x(t)x(t+\tau) \rangle = \frac{k_B T}{m \omega_0^2} e^{\frac{- \gamma}{2 m}t}[\cosh(b t) + \frac{\gamma}{2 m b}\sinh(b t)] 
\end{equation}
where $b  = \frac{1}{2}\sqrt{\Gamma^2 - 4\omega_0^2}$ and $\omega_0=\sqrt{\frac{\kappa}{m}}$ is the natural angular frequency of the trap.

In highly overdamped cases, where $\Gamma^2 \gg 4\omega_0^2$, it is also possible to further approximate Equation (\ref{eq:Langevin2}) by omitting the inertial term \cite{reif2009fundamentals} so that the equation of motion becomes:

\begin{equation}\label{eq:beechapprox}
\gamma \frac{\partial x}{\partial t} + \kappa x = F(t)   
\end{equation}

resulting in a simplified autocorrelation function:

%\begin{equation}
%\langle x(t)x(t+\tau) \rangle = \frac{k_B T}{m \omega_0^2} e^{\frac{-\tau}{\tau_0}} 
%\end{equation}

\begin{equation}\label{eq:beechPAC}
\langle x(t)x(t+\tau) \rangle = \frac{k_B T}{\kappa} e^{\frac{-\tau}{\tau_0}} 
\end{equation}

with a well-defined time constant known as the correlation time $\tau_0 = \frac{\gamma}{\kappa}$.  Reconstructions of the correlation function in Equations (3) and (5) can be used to corroborate the mass measurements obtained through the drop-and-restore experiments discussed in this paper. 

In the first of these experiments, the trapped particle is repeatedly released from the ODF trap.  The motion of the particle falling in gravity is modelled by:
\begin{equation}\label{eq:dDrop}
 \frac{\partial^2 x}{\partial t^2} + \Gamma \frac{\partial x}{\partial t} - g = A(t) 
\end{equation}
where $g = -9.8$~m/s$^2$ is the acceleration due to gravity in this coordinate system. Here, since we average uncorrelated repetitions, the stochastic driving term plays no role and the resulting solution to Equation (\ref{eq:dDrop}) is given by:

\begin{equation}\label{eq:Dropsol}
x(t) = \frac{g}{\Gamma}[t + (\frac{1}{\Gamma}+\frac{v_r}{g})( e^{-\Gamma t}-1)]
\end{equation}
where $v_r$ represents the initial velocity of the released particle, which should average to zero over many uncorrelated repetitions.
Therefore, a fit to the displacement-time graph of a falling particle can be used to extract $\Gamma$.

In the subsequent experiment, when the laser confinement is turned on, the particle is restored to the trap center.  This behavior is modelled by:

\begin{equation}\label{eq:dRest}
 \frac{\partial^2 x}{\partial t^2} + \Gamma \frac{\partial x}{\partial t} + \omega_0^2 x - g  = A(t)   .
\end{equation}

Once again, since numerous uncorrelated restorations are averaged, the stochastic drive does not contribute to the resulting effective solution to Equation (\ref{eq:dRest}), which is given by: 
\begin{equation}\label{eq:Restsol}
x(t) = x_0e^{\frac{-\Gamma}{2}t}[\cosh(b t) + \frac{\Gamma}{2 b}\sinh(b t)]+\frac{v_0}{b}[e^{\frac{-\Gamma}{2}t}\sinh(b t)] 
\end{equation}

where $x_0$ is the initial position and and $v_0$ is the recapture velocity of the particle at the time when the laser force is turned on to restore the particle.  Thus it is possible to infer the value of \textit{m} from a fit to Equation (\ref{eq:Restsol}) using values of $\Gamma$ from the drop experiments, and $\kappa$ from independent measurements of the trap spring constant.

We now comment on the expectations for the recapture velocity in Equation (\ref{eq:Restsol}), where for drop times $t \gg \tau_p$, $v_0$ can be estimated as the sum of the terminal velocity and the effect of the ODF during the first frame of exposure.  The variation in the recapture velocity as a function of drop time can be modelled by

\begin{equation}\label{eq:v0simp}
v_0(t) = v_T - t_{exp}\kappa x(t)/m,
\end{equation}

where $v_T = \frac{g}{\Gamma}$ is the terminal velocity of the particle, $t_{exp}$ is the exposure time for a single frame of acquisition, and $x(t)$ is the trajectory described by Equation (\ref{eq:Dropsol}).

\section{Experimental Details}

\begin{figure}%
\centering
{\includegraphics[width=0.955\linewidth]{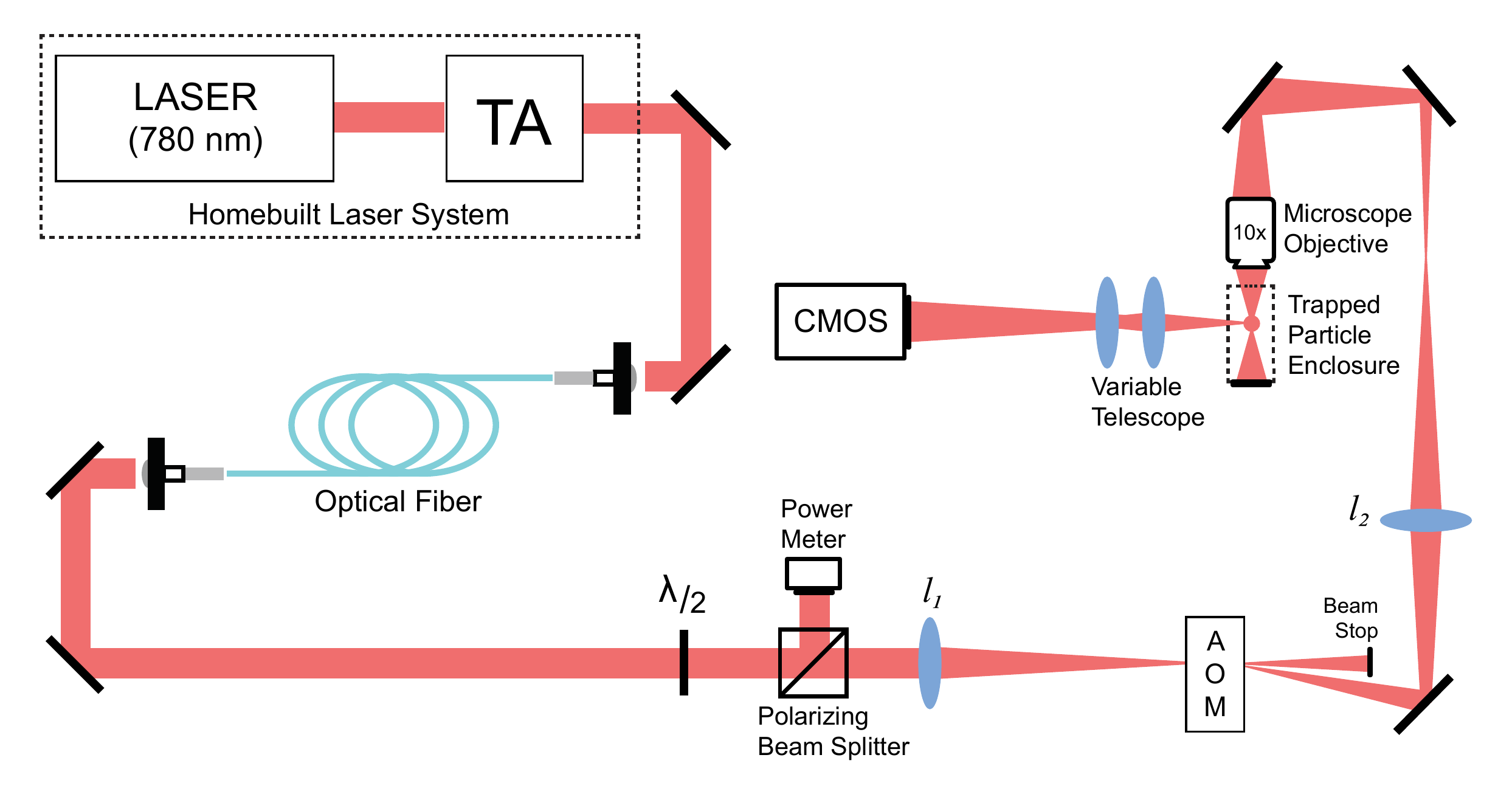}\label{subfig:exp}}
\caption{\label{fig:exp}(Color online) Schematic diagram of the experimental set-up.  The focal lengths of the beam shaping lenses are: $l_1 \sim 45$~cm and $l_2 \sim 30$~cm.  The mirrors in between the AOM and the 10$\times$ objective act as a periscope such that the beam entering the objective is directed downward along the vertical direction. Here, TA represents the tapered amplifier, AOM represents the acousto-optic modulator, and CMOS represents the camera.}
\end{figure}

\begin{figure}
\centering
\subfloat[][]{\includegraphics[width=0.44949\linewidth]{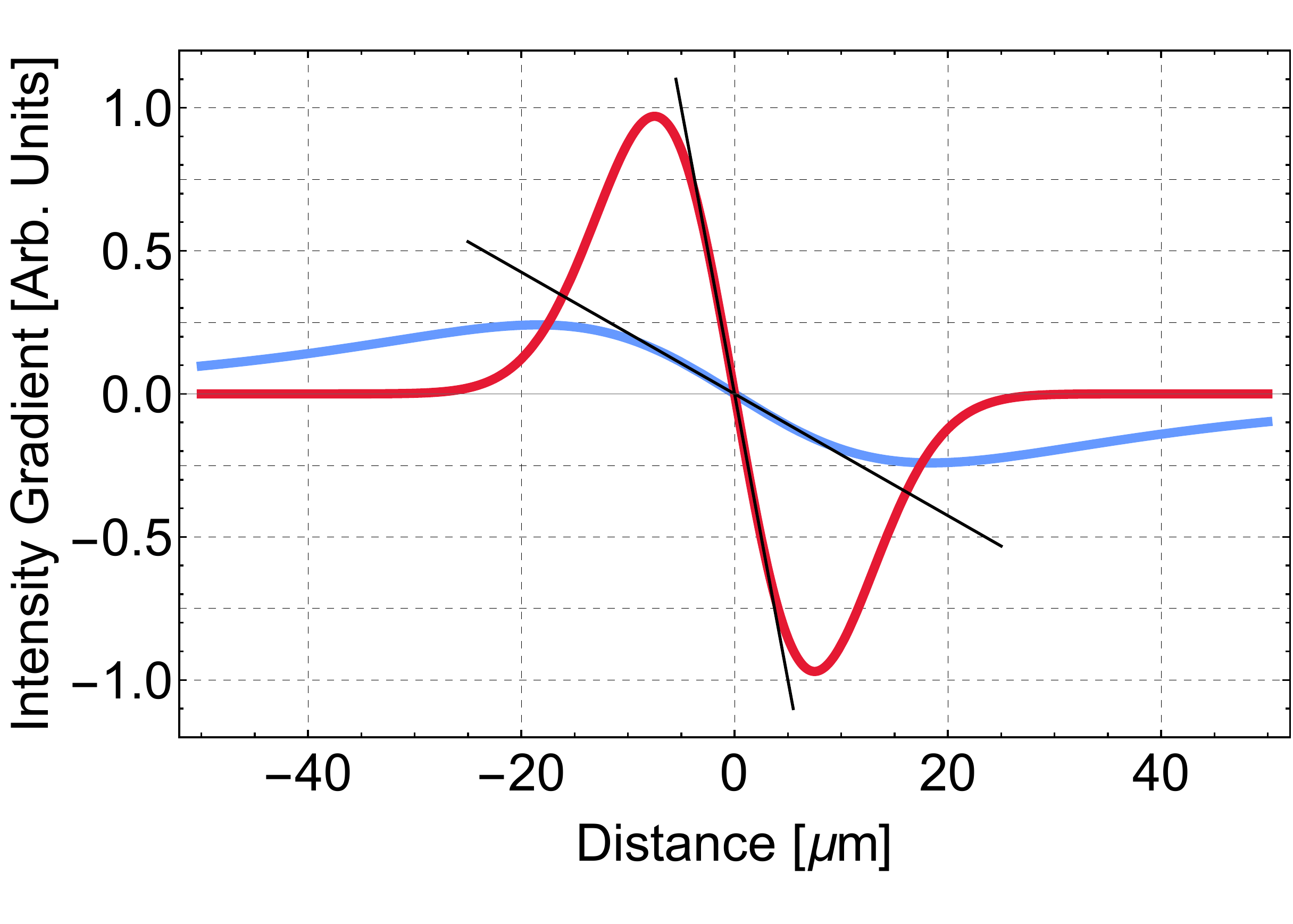}\label{subfig:2grad1}}
   \qquad
\subfloat[][]{\includegraphics[width=0.44949\linewidth]{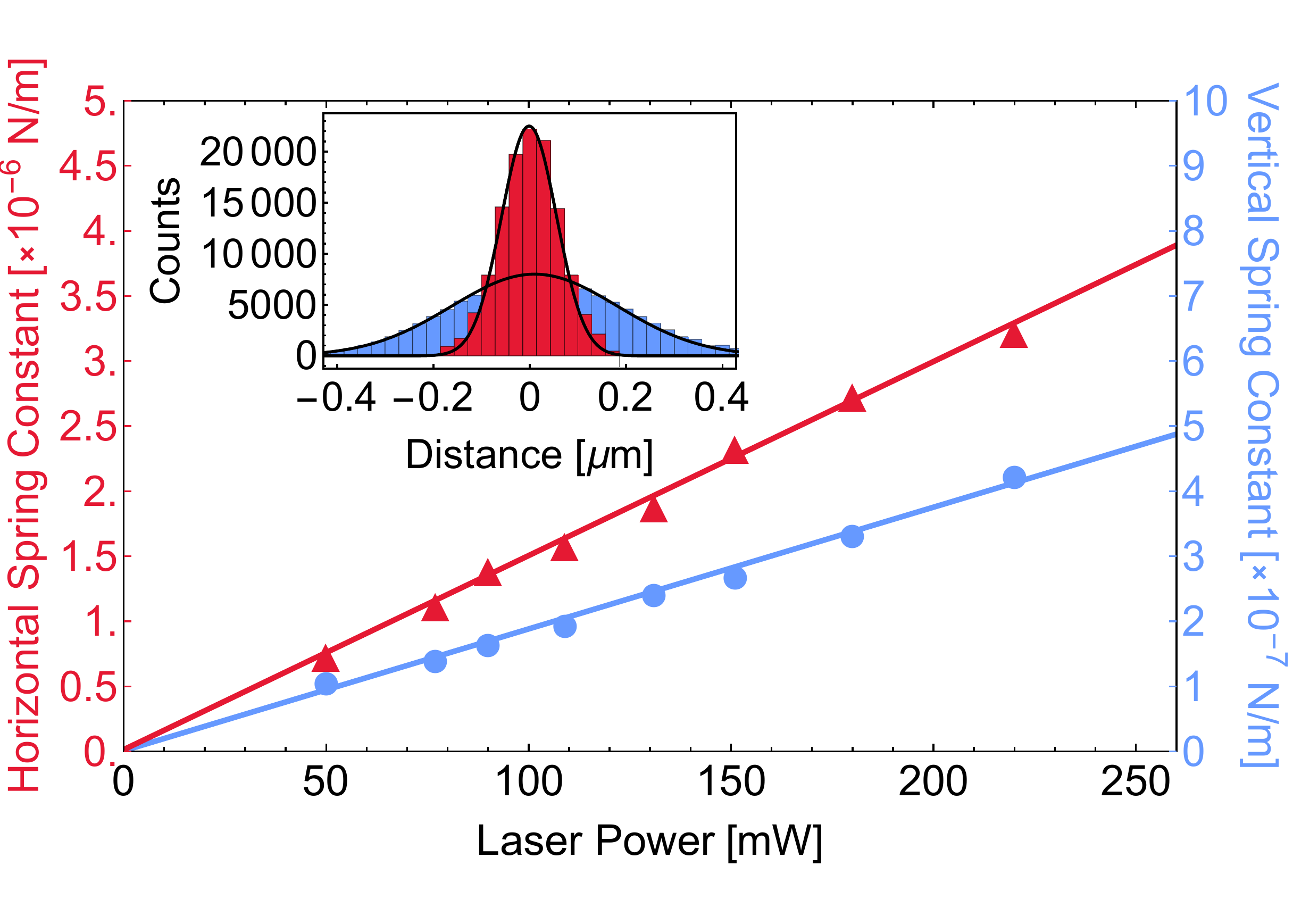}\label{subfig:2k1}}
   \qquad
\caption{\label{fig:2beam}(Color online) (a) Intensity gradients along the vertical (blue/light gray) and horizontal (red/dark gray) directions.  
The straight lines indicate the linear (harmonic) ranges of the potentials.  
 (b) The spring constant as a function of laser power along the horizontal (left-red/dark gray axis) and vertical (right-blue/light gray axis) directions. The linear fits to the two data sets give $\kappa = [(1.49\pm0.04)\times10^{-8}~\frac{\mathrm{N/m}}{\mathrm{mW}}]$ \textit{P} + $[(1.36\pm5.61)\times10^{-8}]$~N/m
along the horizontal direction and $\kappa = [(1.87\pm0.07)\times10^{-9}~\frac{\mathrm{N/m}}{\mathrm{mW}}]$ \textit{P} + $[(1.11\pm4.16)\times10^{-8}]$~N/m, along the vertical, where \textit{P} is the laser power in mW.
Inset shows examples of position histograms in the vertical (blue/light gray) and horizontal (red/dark gray) directions for a representative laser power (77.5 mW).  The black lines show Gaussian fits whose widths are used to calculate the spring constants.}
\end{figure}

The experiments were carried out with a homebuilt laser system, operating at 780 nm, consisting of a master oscillator and semiconductor waveguide tapered amplifier (TA) placed on a pneumatically-isolated optical table.  A schematic of the experimental set-up is shown in Figure \ref{subfig:exp}.  The power stability of the master oscillator has a characteristic Allan deviation of $5\times10^{-6}$ at 10 s~\cite{becica2019RSI} and the TA has an output power of $\sim2$ W \cite{pouliot2018auto}.  The output of the TA was fiber coupled and gently focused through an AOM driven at 80 MHz so that the diffracted beam could be turned off or on in $\sim$150 ns.  As a result, it was possible to rapidly release the trapped particle in a gravitational field, and subsequently restore the particle to its equilibrium position.  The turn-on and turn-off of the diffracted beam from the AOM was controlled by a pulse generator operated at repetition rates ranging from 0.5-20 Hz.  The pulse width that defines the free-fall time of the particle is precise to the level of 1 ns.  
The maximum power in the diffracted beam (250 mW) was controlled with a waveplate and polarizing cube beam splitter.  The diffracted beam was expanded and focused through a 10$\times$ microscope objective (NA 0.25) so that the focus of the beam was $\sim 5 $ mm from the end face of the objective lens.  The intensity gradients associated with the ODF trap were characterized using a scanning knife edge spatial profiler as shown in Figure \ref{subfig:2grad1}.  The focal region was surrounded by a tightly-sealed enclosure with sliding glass windows to reduce air currents.  In this free-space configuration, the trapped particles were introduced by ablating from the tip of a permanent marker inserted into the enclosure.
We note that this is a simple and effective technique for introducing particles into a free-space optical tweezers set-up since the ablated particles have near zero velocity.
Other techniques for introducing trapped particles are described in \cite{ashkin1975optical,li2010measurement,li2013brownian,burnham2010parameter,polster2011trampoline,esseling2012photophoretic}.
The light scattered from the trapped particle in the transverse direction was imaged onto a 
CMOS sensor using a simple two-lens telescope with a variable magnification ranging from $\sim 40\times -~80\times$. 
The CMOS sensor (Phantom UHS-12 v2012) consisted of an 800 $\times$ 1280 pixel array with an overall size of 2.24 cm $\times$ 3.58 cm, which amounts to a pixel size of~28~$\mu$m.  
The camera was operated in continuous mode at a variable frame rate ranging from $1\times 10^4$ to $2\times10^5$ frames per second (fps).
The imaging system was calibrated by photographing a ruled micrometer slide placed in the object plane of the telescope. The calibration involved fitting the profiles of successive rulings in the image plane to Gaussians and determining their separations in pixel units.  Image sequences were stored in on-board memory and transferred to a computer for data processing.
For the drop-and-restore experiments, 100 independent image sequences are averaged to improve the signal-to-noise ratio.  In contrast, the PACF measurements relied on a continuous record length of images.  To compensate for the lack of averaging in the PACF measurements, an intensity filter was used to reduce the effect of broadband background noise entering the telescope.

\section{Results}

\subsection{Spring Constant Determination}

Figure \ref{subfig:2k1} shows the measurement of the spring constant of the ODF trap as a function of laser power.  For each laser power, the spring constant was obtained from a Gaussian fit to the histograms of instantaneous positions (inset in Figure \ref{subfig:2k1}).  The Gaussian fit has a functional form $G(x) = C e^{-\frac{\kappa x^2}{k_BT}}$, where $x$ is the instantaneous position and \textit{C} is a normalization constant \cite{neuman2004optical}.  
Here, the particle positions were recorded with an exposure time of $10~ \mu$s on a suitably long timescale ($t\gg\tau_0$) to ensure uncorrelated measurements.  
This method of determining the trap spring constant is independent of measurements of the damping or the particle mass, contrasting with alternative approaches that rely on the power spectrum.
From linear fits in Figure \ref{subfig:2k1}, we obtain spring constants of $1.49\times10^{-6}$~N/m in the horizontal direction and $1.87\times10^{-7}$~N/m in the vertical direction, for a typical laser power of 100 mW.
We note that the offsets predicted by the fit equations in Figure \ref{subfig:2k1}, which are small, can be used to estimate the inherent noise in the detection system \cite{bechhoefer2002faster}.  
We also note that relative values of the spring constants are consistent with the magnitudes of their respective intensity gradients (see Figure \ref{subfig:2grad1}).

\subsection{Mass Determination from Drop-and-Restore Experiments}

\begin{figure}%
\centering
\subfloat[][]{\includegraphics[width=0.44949\linewidth]{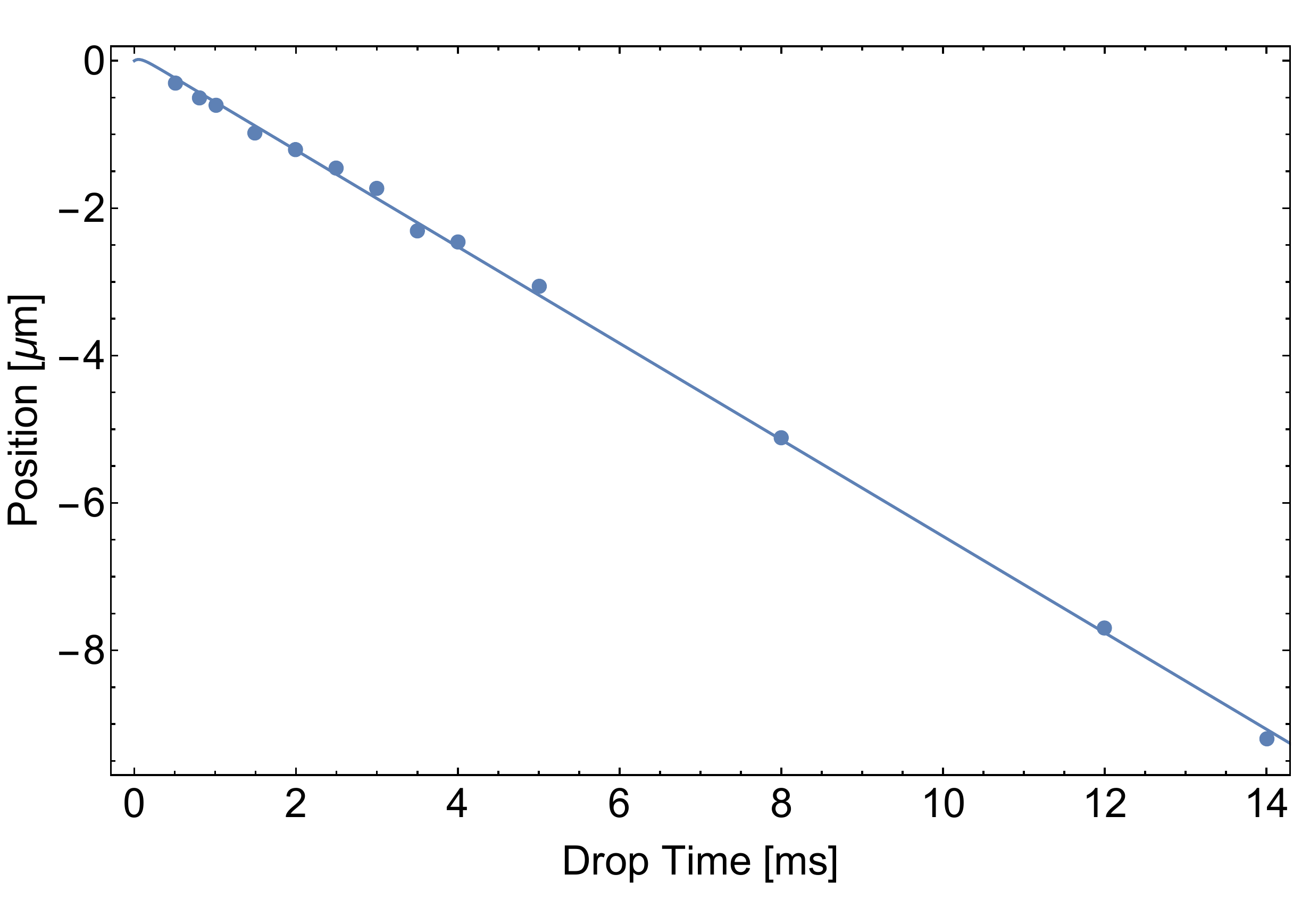}\label{subfig:3drop1}}
   \qquad
\subfloat[][]{\includegraphics[width=0.44949\linewidth]{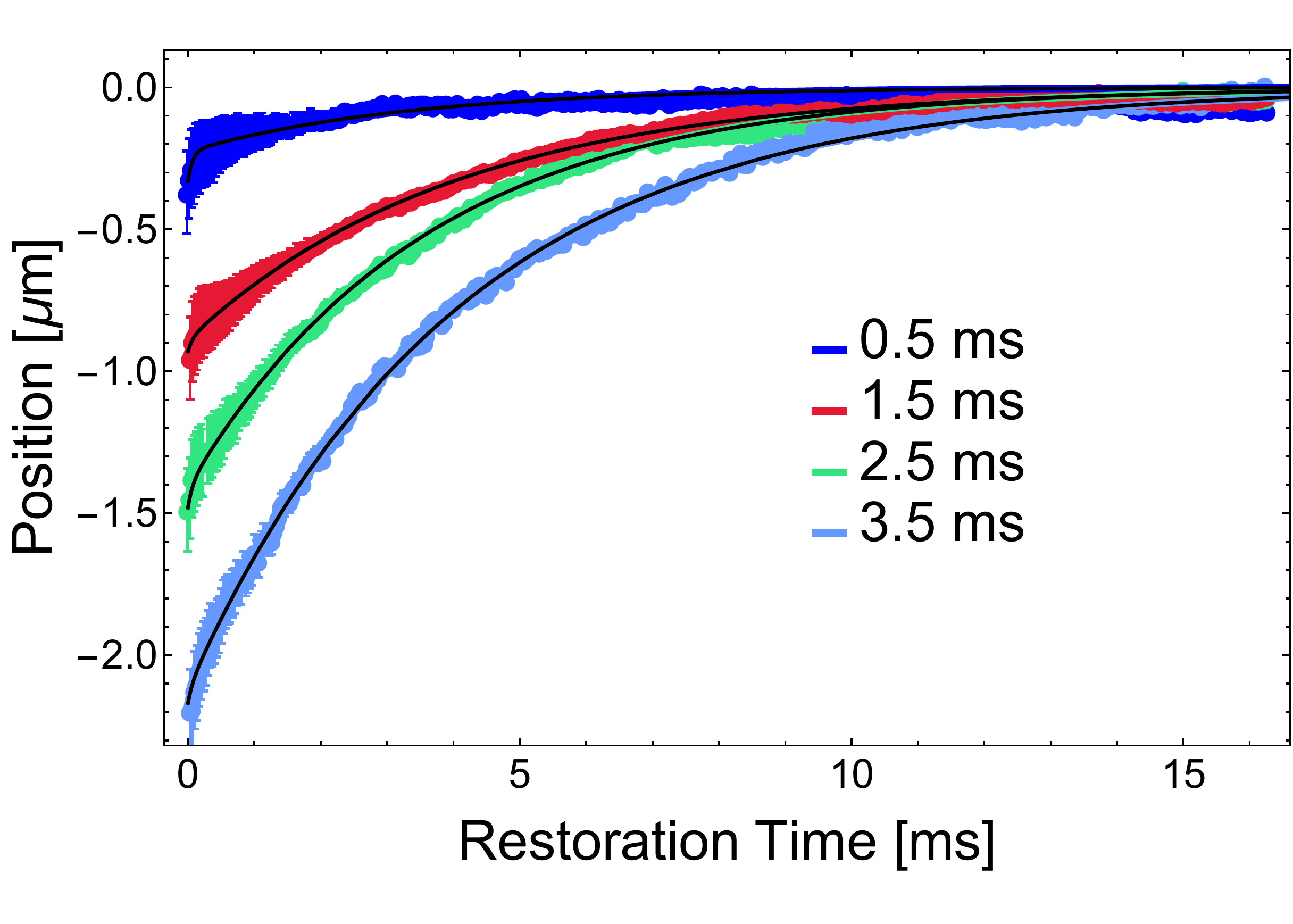}\label{subfig:3restore5}}
\caption{\label{fig:3restore}(Color online) (a) Shows the fall distance as a function of drop time and a fit to Equation (\protect{\ref{eq:Dropsol}}) with $\Gamma = 15.1\pm0.1$ kHz and $v_r = 0.7\pm0.5~\frac{\mu \mathrm{m}}{\mathrm{ms}}$.  (b) Shows the restoration trajectories along the vertical axis of the trapping beam, for a representative set of drop times.  Fits to Equation (\protect{\ref{eq:Restsol}}) are superimposed on the data in black.  The spring constant for these restorations was $\kappa = 2.2\times10^{-7}$~N/m.  The data for both the drop and the restore experiments represent averages of 100 independent repetitions and the error bars indicate the standard deviation of these repetitions.  Here, we take the value of \textit{g} to be -9.80 m/s$^2$.}
\end{figure}

Figure \ref{subfig:3drop1} shows the position of the released particles as a function of ``drop time" (i.e. the time after release from the trap).  The position after each drop time is determined by averaging 100 individual uncorrelated repetitions.  This free-fall data is fit to Equation (\ref{eq:Dropsol}) to determine $\Gamma$, with a statistical uncertainty of $\sim$1\%.  Since the system is highly damped, the trajectory is dominated by the linear term in Equation (\ref{eq:Dropsol}), the slope of which defines $v_T$.

Figure \ref{subfig:3restore5} shows representative trajectories of particles that are being restored to the equilibrium position of the trap, after various drop times.  
The overall data collection time for a set of 13 drop-and-restore experiments was $\sim 90$ seconds.  The restoration trajectories are fit to Equation (\ref{eq:Restsol}) on the basis of known values for $\kappa$ from the calibration (see Figure \ref{subfig:2k1}), as well as $\Gamma$ and $x_0$ from the drop experiment (see Figure \ref{subfig:3drop1}).
Therefore, we are able to determine the mass of the falling particle from a two-parameter fit involving $m$ and $v_0$.  

\begin{figure}%
\centering
\subfloat[][]{\includegraphics[width=0.44949\linewidth]{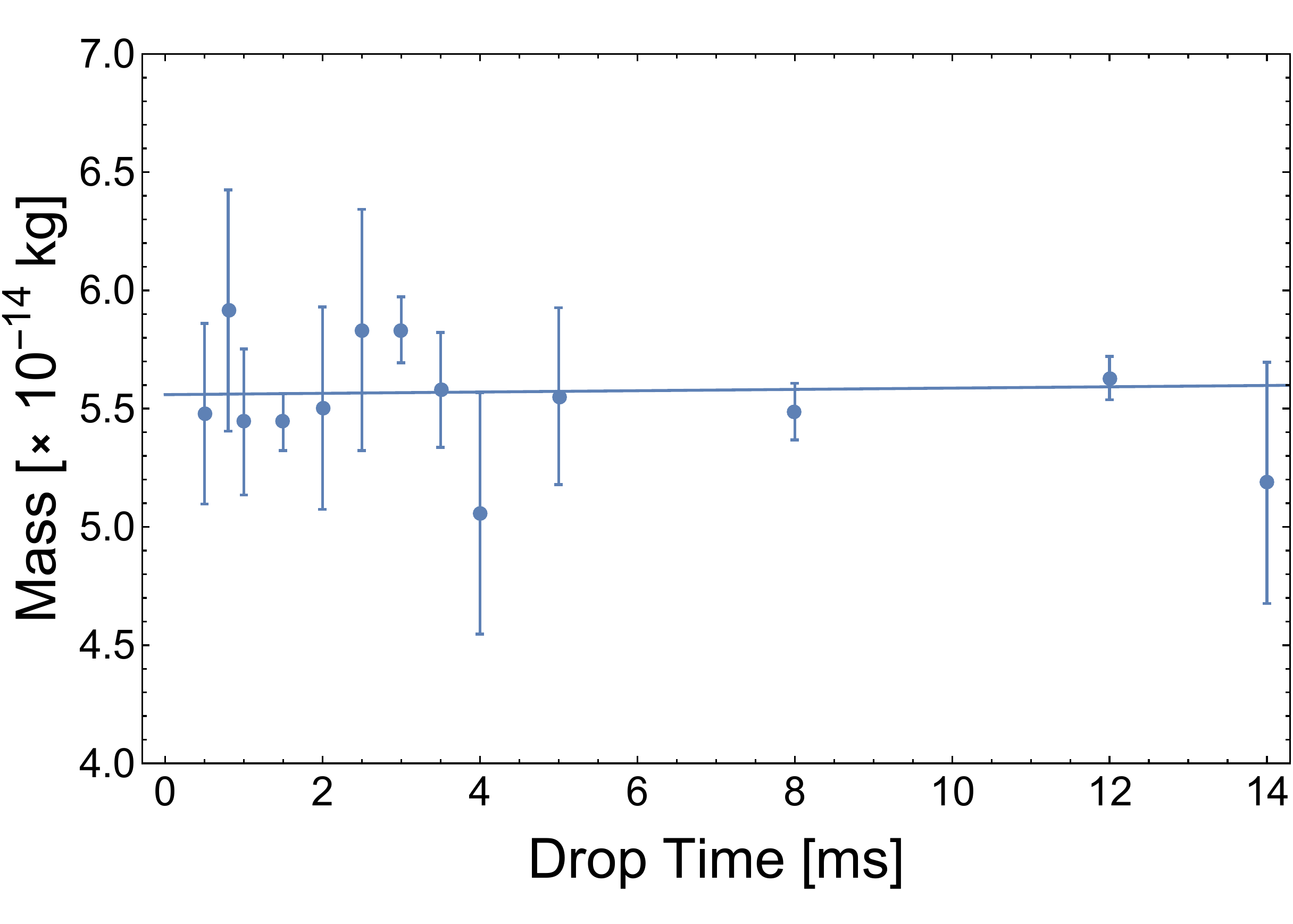}\label{subfig:mass13}}
   \qquad
\subfloat[][]{\includegraphics[width=0.48478749\linewidth]{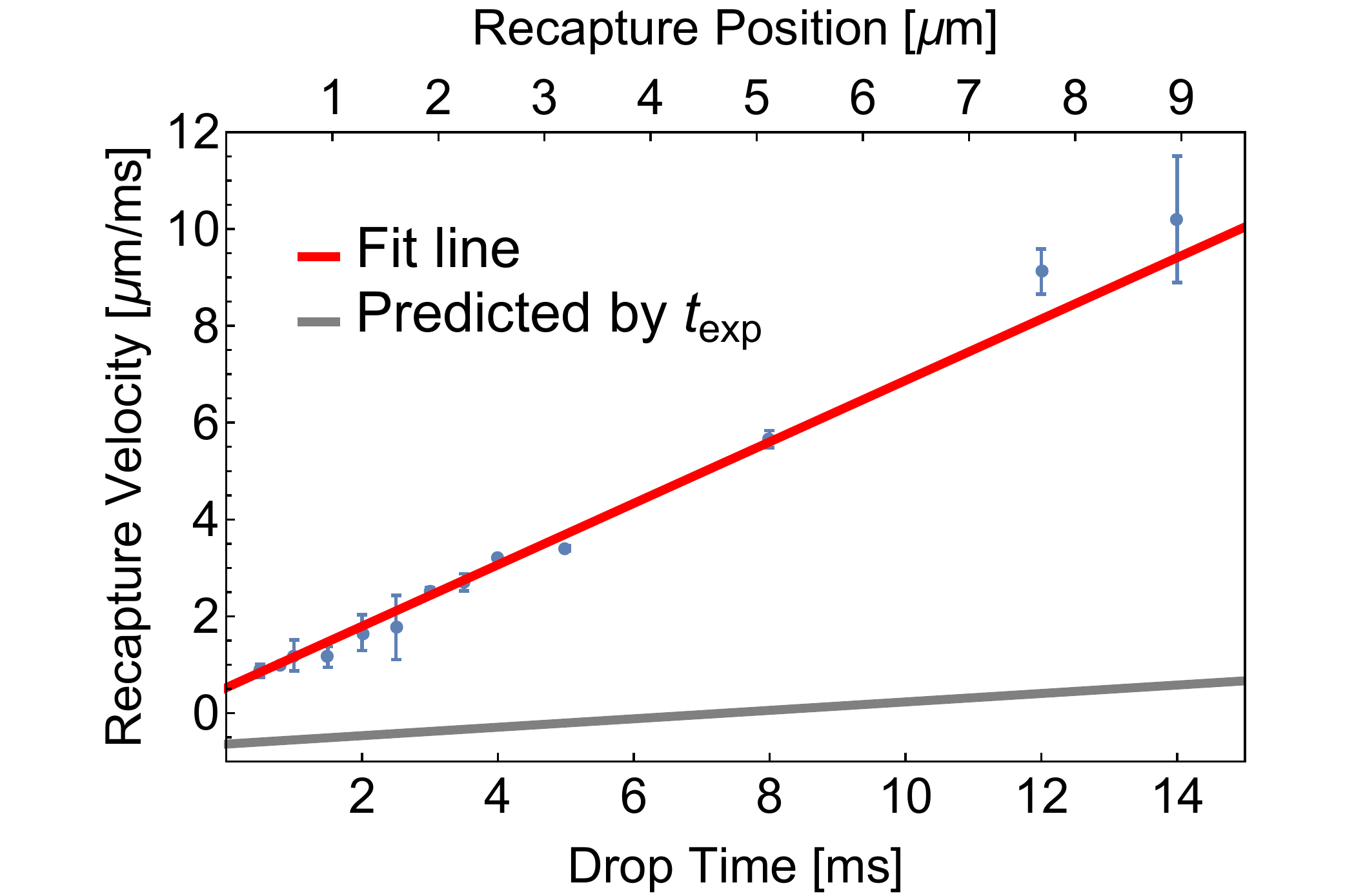}\label{subfig:v0}}
\caption{\label{fig:5masses}(Color online) (a) Mass determined using the drop-and-restore technique for various drop times.  The restoration fits were performed using the values $\Gamma = 1.511\times10^{4}$~Hz from the drop experiment and $\kappa = 2.2\times10^{-7}$ N/m from the vertical spring constant measurements.  
Here the fit line gives an offset value of $(5.58\pm0.08)\times10^{-14}$ kg and a slope which is consistent with zero as expected, namely, $(2.74\pm9.86)\times10^{-17}$~kg/ms.
(b) Fit values of $v_0$ from Equation (\protect{\ref{eq:Restsol}}) as a function of drop time (lower axis) and recapture position with respect to the trap center (upper axis).  The red fit line as a function of the drop time gives an offset value of $v_0(t=0) = (0.5\pm0.1) \frac{\mu\mathrm{m}}{\mathrm{ms}}$, and an acceleration given by the slope of $(0.63\pm0.03) \frac{\mu\mathrm{m}}{\mathrm{ms}^2}$.
The predicted value of the recapture velocity, also as a function of drop time, as defined by Equation (\protect{\ref{eq:v0simp}}), is shown by the gray trendline.
}
\end{figure}

Figure \ref{subfig:mass13} shows the mass extracted from the restoration trajectories for the drop times shown in Figure \ref{subfig:3drop1}.  We find no systematic dependence on the drop time.  
The error bar represents the statistical uncertainty of the single parameter fit.  
We report a mass measurement of $5.58\times 10^{-14}$ kg, with a statistical uncertainty of 1.4\%.
%--------------------
We estimate the overall uncertainty in \textit{m} by numerically varying the parameters $\kappa, \Gamma$, and $x_0$ within experimental error, finding the statistical variation in \textit{m} from the resulting trajectory fits, and combining these individual uncertainties in quadrature.
In this manner, we infer a systematic uncertainty in $m$ of $6\times10^{-15}$ kg ($\sim13\%$).  
%--------------------------

%%%%
Figure \ref{subfig:v0} shows the fit values of the recapture velocity $v_0$, as a function of the drop time and recapture position, for each of the mass determinations shown in Figure \ref{subfig:mass13}.
The red fit line shows that the initial recapture velocity continues to increase as the particle is allowed to fall further from the equilibrium position.
Figure \ref{subfig:v0} also shows the predicted value of the recapture velocity, as defined by Equation (\ref{eq:v0simp}) (gray line).
We attribute the difference between the two trend lines to an impulse proportional to the distance from the trap center imparted by the turn-on and -off of the AOM that produces a transient, uneven illumination of the particle.
Our conjecture is supported by the drop experiments shown in Figure \ref{subfig:3drop1} where the fit to Equation (\ref{eq:Dropsol}) yields a small initial velocity.
We note that this effect, indicative of a small impulse in the drop data, is consistent with the offset extracted from the fit in Figure \ref{subfig:v0}.
We suggest that these features of the data arise because the resultant impulse imparted scales with distance from the beam focus due to the position dependent nature of the ODF.
During the turn-off, or during the turn-on following a short drop time, the position of the particle is near the uniformly illuminated region around the equilibrium position of the trap.
In contrast, when the AOM is turned on after longer drop times, the particle is at increasing distances from the trap center where any uneven and transient illumination due to the AOM will have a larger effect. 
Therefore, the linear dependence of the recapture velocity on the drop time in Figure \ref{subfig:v0} can be attributed to the combined effects of the laser force, the impulse from the AOM, and gravity.
%%
%%%%%

\subsection{Mass Determination from Autocorrelation Functions}

\begin{figure}%
\centering
   \subfloat[][]{\includegraphics[width=0.44949\linewidth]{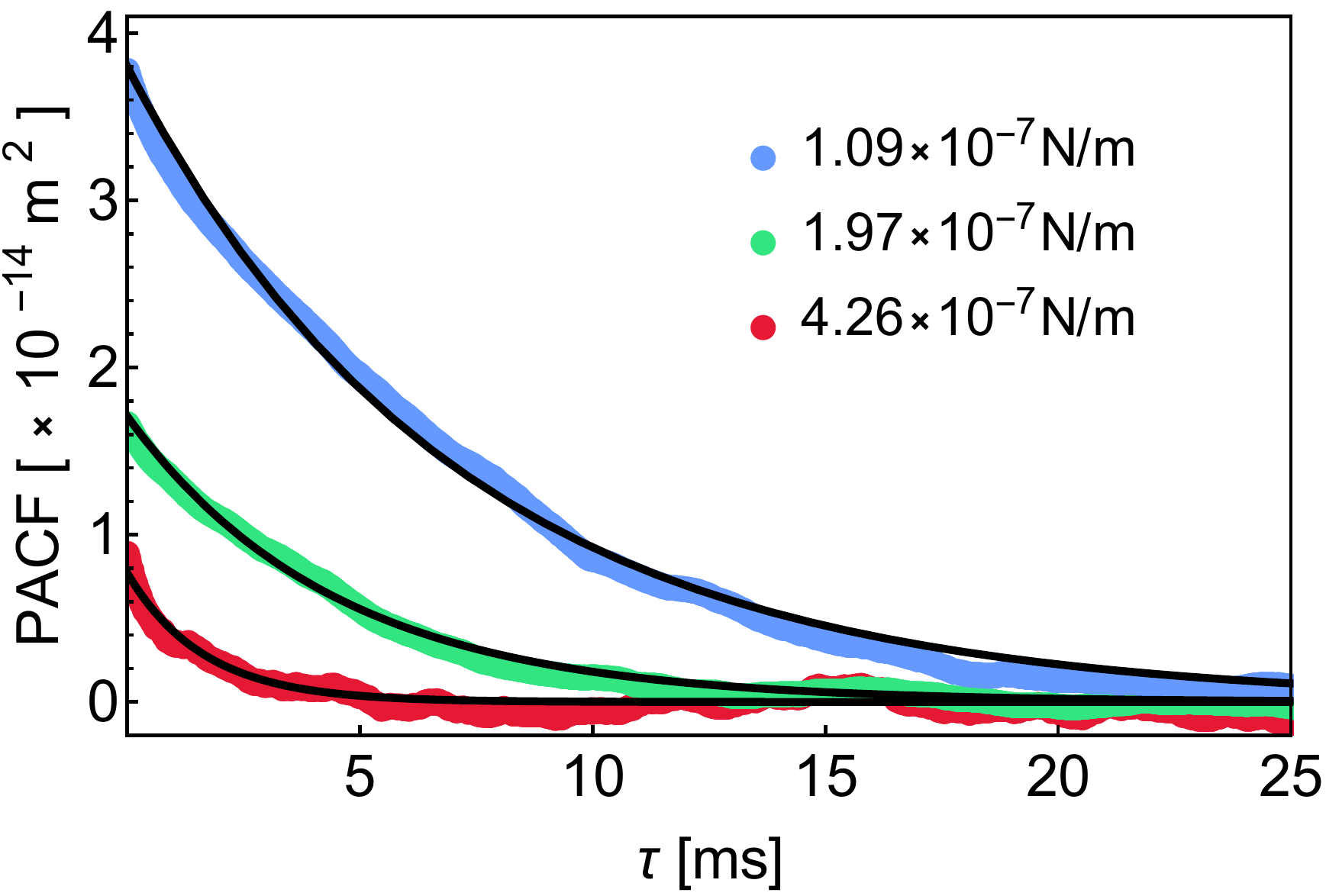}\label{subfig:7pacgam}}
   \qquad
\subfloat[][]{\includegraphics[width=0.44949\linewidth]{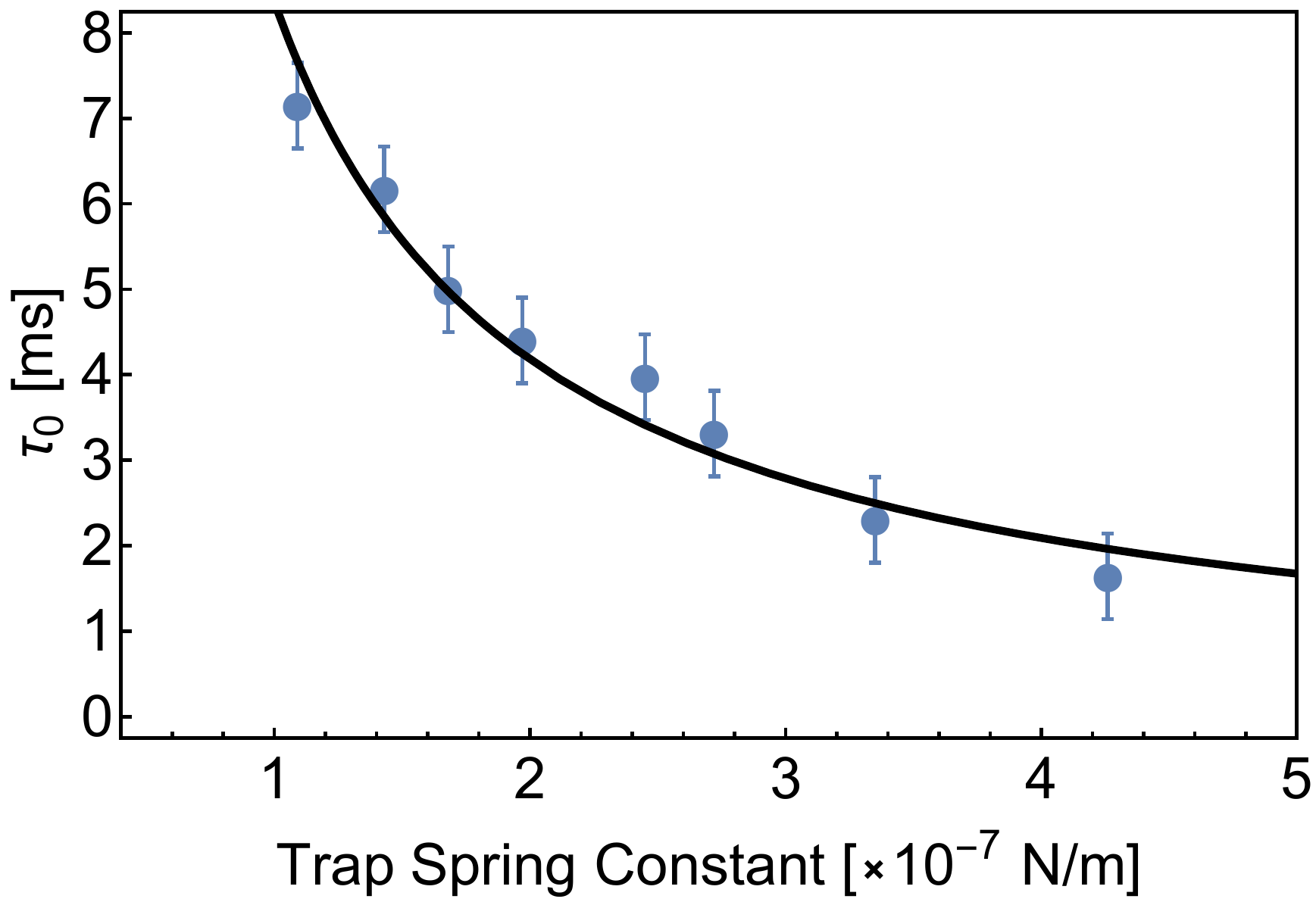}\label{subfig:7tau0}}
\caption{\label{fig:7PACF}(Color online) (a) PACF of particle motion at various laser powers.  The black lines show fits to Equation (\protect{\ref{eq:beechPAC}}), based on the overdamped approximation. Part (b) shows the resulting $\tau_0$ from the PACF fits for a range of trap spring constants.  The fit function is of the form $\tau_0=\gamma/\kappa$, from which we obtain $\gamma=(8.38\pm0.23)\times10^{-10}$ kg/s.}
\end{figure}

Figure \ref{subfig:7pacgam} shows representative examples of PACFs generated from data sets that are several seconds in duration with a frame rate of $10^5$ fps.  This data, obtained at various laser powers, represents the time-domain analog of other techniques for mass determination that rely on the power spectral density \cite{nanoricci2019accurate,lin2017measurement}.
Here, however, the smoothness of the PACF suffers due to the record length, which was restricted to match that of the drop-and-restore experiments.
While the PACFs can be fit to Equation (\ref{eq:fullPACF}), the complex functional form results in an over-estimation of the uncertainty in the mass.  
The large uncertainty persists even if the values of $\Gamma$ and $\kappa$ are constrained on the basis of independent experiments.
As a result, we have used the autocorrelation function in the large damping limit given by Equation (\ref{eq:beechPAC}) to fit the data since it has a much simpler functional form.
From these fits we extract the correlation time constants with a precision of approximately 3\%.

Figure \ref{subfig:7tau0} shows the resulting fit values for the correlation time constant $\tau_0=\gamma/\kappa$, as a function of trap spring constant (which is varied by adjusting the laser power).
The error bars displayed in this figure represent the total uncertainty due to the intensity filter used to reduce the background noise in the PACFs and the inherent uncertainty in the exponential fits.
This data, which exhibits the predicted inverse power dependence, can be used to extract a damping coefficient  $\gamma=(8.38\pm0.23)\times10^{-10}$ kg/s. 
Combining this result with the damping rate $\Gamma$ measured in the drop experiments, we find a mass value of $(5.55\pm0.15)\times10^{-14}$ kg, which corroborates the determination from the trap restoration experiments discussed earlier (see Table \ref{tab:fullsummary}).
If we consider the damping coefficient extracted from Figure \ref{subfig:7tau0} and assume Stokes' law, we find the particle radius to be $(2.3\pm 0.1)~\mu$m, which is comparable to the radius inferred from the images ($2.4 \pm 0.3 ~\mu$m).  We note that this comparison, which is also shown in Table \ref{tab:fullsummary}, takes into account the effects of calibration uncertainties such as absolute resolution, motional blurring, and depth of field.
By combining this radius with the mass, we infer a particle density of ($1.1\pm0.1$)$\times10^3$ kg/m$^3$, which is consistent with the density of resins used in common permanent markers \cite{markers}.

\begin{table*}
\caption{\label{tab:fullsummary}Summary of mass and particle radius measurements based on various techniques.  $^*$Mass determined by combining $\tau_0$ from Figure \protect{\ref{subfig:7tau0}} and $\Gamma$ measured in drop experiments (Figure \protect{\ref{subfig:3drop1})}.  $^{**}$Radius measurements inferred from Stokes' law using the PACF time constants from Figure \protect{\ref{subfig:7tau0}}.}
%\begin{ruledtabular}
\begin{tabular*}{\textwidth}{c@{\extracolsep{\fill}}c c|cc}
 \multicolumn{3}{c}{Mass determination}&\multicolumn{2}{c}{Particle size measurement}\\
 Technique& Mass (kg)&& Technique &Radius ($\mu$m)\\ \hline
Drop-and-restore & $(5.58\pm0.08)\times10^{-14}$ && Direct Observation & $2.4\pm0.3$ \\ 
PACF \& Drop$^*$ & $(5.55\pm0.16)\times10^{-14}$ && PACF \& Stokes$^{**}$& $2.3 \pm 0.1$ \\
\end{tabular*}
%\end{ruledtabular}
\end{table*}

\section{Conclusions}
We have presented a simple and effective technique based on drop-and-restore experiments in a gravitational field to determine the masses of particles confined using free-space optical tweezers. The mass determination, which has a statistical uncertainty of $< 2\%$, has also been corroborated by position autocorrelation measurements. In contrast with other techniques (see Table \ref{tab:masssummary}), our experiments do not require the use of secondary lasers, feedback systems, or vacuum environments.  Instead, our measurements rely on direct imaging of scattered light with a fast CMOS sensor and a straightforward spatial calibration procedure.

We anticipate that the precision of this technique can be further improved by using higher laser powers and a larger Rayleigh range for the focused beam. This combination will increase the recapture range, defined by the turning points of the axial intensity gradient and allow the available field of view to be fully exploited. However, potential complications may arise from heating and local changes in the viscosity of the medium, which should be accounted for at higher laser intensities \cite{bera2016simultaneous,peterman2003laser,mcgloin2008optical}.
Additionally, we expect that the impulses attributed to the AOM turn-on can be significantly suppressed by employing a dual-pass AOM \cite{spirou2003high}.
It is also possible to further reduce the estimated systematic uncertainty by using faster frame rates to improve instantaneous position measurements.  Similarly, the accuracy of spring constant measurements can be improved by actively stabilizing the power output of the AOM using an RF feedback loop and by using temperature-insensitive polarizers.

Our drop-and-restore method may also be used to study highly absorbing particles confined in photophoretic traps provided the effects of amplitude modulation in such traps are carefully modeled \cite{chen2018drop,lin2017measurement}. Other extensions could involve the investigation of particulates trapped in liquids or media of higher viscosity. Based on the statistical precision, we expect that this technique should be applicable to the discrimination of contaminants in flue gases as well as biological agents such as pollen and pathogens trapped in free space and liquid cultures \cite{ashkin1987bacteria,pang2014optical,lin2015pulling,zhang2019newbac}. Given the data acquisition time of approximately 90 s, we anticipate that this work will open the door for the rapid determination of relative masses of a variety of trapped particles in future studies.

\section{Acknowledgements}
This work is supported by: Canada Foundation for Innovation, Ontario Innovation Trust, Ontario Centers of Excellence, Natural Sciences and Engineering Research Council of Canada, and York University.

We thank Chris Wernick of Delta Photonics for a week-long loan of a high-speed camera.  We also acknowledge helpful discussions with Matthew George and Ozzy Mermut.

\providecommand{\noopsort}[1]{}\providecommand{\singleletter}[1]{#1}%

\end{document}